\documentclass[10pt, conference, twocolumn]{IEEEtran}

\usepackage{graphics}
\usepackage{amsmath}
\usepackage{amssymb}
\usepackage{graphicx}

\newtheorem{theorem}{\bf Theorem}[section]

\newcommand{\m}[1]{\mathbf{#1}_1^m}
\newcommand{\lo}[1]{\log_2\left(#1\right)}
\newcommand{\lon}[1]{\ln\left(#1\right)}
\newcommand{\mk}[1]{\mathbf{#1}_1^k}

\newcommand{\pe}{{\langle P_e\rangle}}

\title{Green Codes: Energy-Efficient Short-Range Communication}
\author{Pulkit Grover and Anant Sahai\\ Wireless Foundations, Department of EECS
\\University of California at Berkeley, CA-94720, USA\\\{pulkit,  sahai\}@eecs.berkeley.edu}

\begin{document}

\maketitle

\begin{abstract}

   A green code attempts to minimize the total energy per-bit required
   to communicate across a noisy channel. The classical
   information-theoretic approach neglects the energy expended in
   processing the data at the encoder and the decoder and only
   minimizes the energy required for transmissions. Since there is no
   cost associated with using more degrees of freedom, the
   traditionally optimal strategy is to communicate at rate zero.

   In this work, we use our recently proposed model for the power 
consumed by iterative message passing. Using
   generalized sphere-packing bounds on the decoding power, we find
   lower bounds on the total energy consumed in the transmissions and
   the decoding, allowing for freedom in the choice of the rate. We show
   that contrary to the classical intuition, the rate
   for green codes is bounded away from zero for any given error
   probability. In fact, as the desired bit-error probability goes to
   zero, the optimizing rate for our bounds converges to $1$.
\end{abstract}

\section{Introduction}
With the development of billion transistor chips, the range of communication has come down dramatically from hundreds of kilometers (e.g. deep space communication) to a few meters (e.g. ad-hoc wireless networks) or a few millimeters or even less (e.g. on chip communication). To communicate over smaller distances, the transmit power required is much smaller. At these distances, the energy used in transmissions can be comparable to that expended by the system processes. The small size limits the ability of these chips to dissipate heat. Further, the chip might be battery operated, imposing stringent constraints on its energy usage. It is therefore of interest to design coding techniques that minimize the \textit{total energy} consumed, which includes the transmission energy as well as the processing energy. We refer to the coding techniques that minimize the total energy as \textit{green codes}.

The classical information theoretic approach finds the minimum \textit{transmission} energy required to communicate reliably across the channel. The approach is motivated by long-range communication, that corresponds to power constrained channels. Shannon~\cite{ShannonOriginalPaper} first characterized the minimum energy required to communicate across a channel with fixed rate. The resulting bounds are expressed using `waterfall' curves that convey the revolutionary idea that unboundedly low probabilities of bit-error are attainable using only finite transmit power. 
This characterization raises a natural question: what is the minimum energy required for communication that is free of a rate constraint? The classical approach~\cite{pierce}~\cite{verducapacitycost} gives the minimum transmission energy required (on average) to communicate one bit reliably across the channel. For example, for an AWGN channel of noise variance 1, this minimum energy is 
\begin{equation}
\lim_{P_T\rightarrow 0}\frac{P_T}{C(P_T)}=2\ln(2)\;\text{ Joules}.
\end{equation}
Since there is no penalty associated with lower rates, it is good to use as many degrees of freedom as are available, and the optimal transmission rate is zero.

The problem of minimizing combined transmission and processing energy is well studied in networks. The common thread in \cite{pagrawal,goldsmithbahai,goldsmithwicker, massaad, Vasudevan,kravertz} is that the energy consumed in processing the signals can be a substantial fraction of the total power.
In~\cite{massaad}, an information-theoretic
formulation is considered. The authors model the processing energy by a constant $\epsilon$ per unit time when the transmitter is transmitting (and hence, is in the `on' state). A total of $r$ channel uses are allowed, and the total energy available is $r\mathcal{E}$, where $\mathcal{E}$ is a constant. Let $P_i$ be the transmit power at $i$-th time instant, and let $C(P_i)$ be the capacity of the corresponding channel. Then the problem is to transmit maximum number of bits with the total power less than $r \mathcal{E}$. That is, 
\begin{eqnarray}
&\max &\sum_{i=1}^r 1_{i}C(P_i)\\
&\text{subject to}& \;\;\sum_{i=1}^r 1_i  (P_i+\epsilon)\leq r\mathcal{E}
\end{eqnarray}
where $1_i= 1$ if a symbol is transmitted in the $i$-th channel use, and is 0 otherwise. This is equivalent to dividing the channel into $r$ sub-channels, with independent coding on each sub-channel. Since the capacity function $C(P)$ is concave in its argument, for maximizing the total number of information bits communicated, the transmission energy $P_i$ should be equal for all $i$ where $1_i=1$. Without accounting for the energy consumed by the system processes, the optimal strategy would be to use all the $r$ parallel channels, and share the energy equally amongst them. However, the energy consumed by the system processes imposes a fixed penalty on each channel use. The authors quantify this tension by measuring `burstiness'  $\Theta$ of signaling defined as $\Theta = \frac{1}{r}\sum_{i=1}^r 1_{i}$.

The transmissions should not be too bursty because of the law of diminishing returns associated with the $\log(\cdot{})$ function. On the other hand, the transmission strategy should not make use of all degrees of freedom either, since there is an $\epsilon$ cost associated with the use of each degree of freedom. The authors conclude that for minimum total energy, $0<\Theta<1$.  Contrary to conventional information theoretic wisdom, it is no longer optimal to use all available degrees of freedom. Consequently, the optimal rate that minimizes the total energy consumption is bounded away from zero. That is, \textit{if processing energy is taken into account, green codes may not communicate at zero rate!}

The objective in~\cite{massaad}~\cite{goldsmithbahai}~\cite{kravertz} is to reduce the energy consumption for wireless devices that consume energy continuously when operating e.g. hand-held computers, high-end laptops, etc. Energy consumption per unit time for such devices is indeed well modeled by a constant possibly independent of the coding strategy being used. In this paper, we are interested in the energy expended by the decoding process itself. The decoding circuit requires some non-zero energy to perform each operation. As opposed to energy consumed by system processes in~\cite{massaad,goldsmithbahai,kravertz}, the decoding energy depends significantly on the code construction, the rate and the desired error probability, and therefore needs more careful modeling.


In this work, we study explicit models of energy expended at the decoder. Owing to their low implementation complexity, and hence low energy consumption, we concentrate on the message passing decoder. For this decoder, we derive lower bounds on the combined transmission and decoding energy, with no constraint on the rate. We show that the optimizing rate for green codes based on message passing decoding is indeed bounded away from zero. As the error probability decreases to zero, the optimizing rate increases. In a result that is qualitatively different from those in~\cite{massaad}, we show that there is no advantage in increasing the rate beyond $1$. Therefore, as the error probability converges to zero, the optimizing rate converges to 1! 

The organization of the paper is as follows : In Section~\ref{sec:model}, we introduce the channel model, the decoder model, and the energy model. In Section~\ref{sec:summary}, we summarize some of our  results in~\cite{waterslide}. In Section~\ref{sec:minenergy}, we build on the results in~\cite{waterslide} to find bounds on the minimum total energy required to communicate across a channel, with no rate constraint, taking into account the decoding energy as well. We conclude in Section~\ref{sec:conclusions}.

\section{System model}
\label{sec:model}
 Consider a point-to-point communication
link. An information sequence $\mk{B}$ is encoded into $2^{mR}$ codeword $\m{X}$, using a possibly randomized encoder.  The observed
channel output is $\m{Y}$. The information sequences are assumed to
consist of iid fair coin tosses and hence the rate of the code is
$R=k/m$. 

The channel model considered is an average power constrained AWGN channel of noise variance $\sigma_\mathcal{P}^2$. We also obtain some results for the BSC arising from performing hard-decision on BPSK symbols transmitted over an AWGN channel. The true channel is denoted by $\mathcal{P}$. The channel capacity is denoted by $C_{\sigma^2}(P_T)$, where $\sigma^2$ is the noise variance, and $P_T$ is the average power constraint. We drop $\sigma^2$ from this notation when no ambiguity is created in doing so.  

For maximum generality, we do not impose any {\em a priori} structure
on the code itself. Instead, inspired by \cite{TarokhVLSI, Shanbhag,
  HotChannel1}, we focus on the parallelism of the
decoder and the energy consumed within it. We assume that the decoder
is physically made of computational nodes that pass messages to each
other in parallel along physical (and hence unchanging) wires. A
subset of nodes are designated `message nodes' in that each is
responsible for decoding the value of a particular message bit.
Another subset of nodes (not necessarily disjoint), called the `observation nodes' has members that
are each initialized with at most one observation of the received
channel output symbols. There may be additional computational nodes to merely help in decoding.

The implementation technology is assumed to dictate that each
computational node is connected to at most $\alpha+1 > 2$ other
nodes\footnote{In practice, this limit could come from the number of
  metal layers on a chip. $\alpha = 1$ would just correspond to a big
  ring of nodes and is therefore uninteresting.} with
bidirectional wires. No other restriction is assumed on the topology
of the decoder. In each iteration, each node sends (possibly
different) messages to all its neighboring nodes. {\bf No restriction
  is placed on the size or content of these messages except for the
  fact that they must depend only on the information that has reached the
  computational node in previous iterations.} If a node wants to
communicate with a more distant node, it has to have its message
relayed through other nodes. The neighborhood size at the end of $l$
iterations is denoted by $n \leq \alpha^{l+1}$. Each computational node is assumed to consume a fixed $E_{node}$
joules of energy at each iteration.

Let the average probability of bit error of a
code be denoted by $\pe$ when it is used over channel $\mathcal{P}$. The main tool 
is a lower bound on the neighborhood size $n$ as a function
of $\pe$ and $R$. This then translates into a lower bound on the
number of iterations that can in turn be used to lower bound the
required decoding power. 

Throughout this paper, we allow the encoding and decoding to be
randomized with all computational nodes allowed to share a pool
of common randomness. We use the term `average probability of error'
to refer to the probability of bit error averaged over the channel
realizations, the messages, the encoding, and the decoding.

\section{Lower bounds on the decoding complexity and total energy}
\label{sec:summary}
In this section we summarize our results for lower bounds on decoding complexity for an AWGN channel from~\cite{waterslide}. 
The main bounds are given by theorems that capture a local sphere-packing effect. These can be turned around to give a family of lower bounds on the neighborhood size $n$ as a function of $\pe$ and $R$. Using a simple lower bound on the number of iterations, $l\geq \frac{\log(n)}{\log(\alpha)}-1$, we get a lower bound\footnote{We approximate this by $l\geq \frac{\log(n)}{\log(\alpha)}$ for the rest of the paper.} on complexity. The family of lower bounds is indexed by the
choice of a hypothetical channel $\mathcal{G}$ and the bounds can be optimized numerically for any desired set of parameters.
\begin{theorem}\label{thm:basicAWGNbound}
For the AWGN channel and the decoder model in Section~\ref{sec:model}, let $n$ be the maximum size of the decoding neighborhood of any individual message bit. The following lower bound holds on the average probability
of bit error.
\begin{eqnarray}
\label{eq:peip}
\nonumber
\pe\geq \sup_{\sigma_\mathcal{G}^2:C_{\sigma_\mathcal{G}^2}(P_T)<R}\frac{h_b^{-1}(\delta(\sigma_\mathcal{G}^2))}{2}\exp\bigg{(} -nD(\sigma_\mathcal{G}^2||\sigma_\mathcal{P}^2)\\-\sqrt{n}\left(\frac{3}{2}+2\ln\left(\frac{2}{h_b^{-1}(\delta(\sigma_\mathcal{G}^2))}\right)\right)\left(\frac{\sigma_\mathcal{G}^2}{\sigma_\mathcal{P}^2}-1  \right)   \bigg{)},
\end{eqnarray}
where $\delta(\sigma_\mathcal{G}^2) = 1 -C_{\sigma_\mathcal{G}^2}(P_T)/R$, the capacity $C_{\sigma_\mathcal{G}^2}(P_T) = \frac{1}{2}\log_2\left(1+\frac{P_T}{\sigma_\mathcal{G}^2}\right)$, and the KL divergence $D(\sigma_\mathcal{G}^2||\sigma_\mathcal{P}^2)=\frac{1}{2}\left[ \frac{\sigma_\mathcal{G}^2}{\sigma_\mathcal{P}^2}-1-\ln\left(\frac{\sigma_\mathcal{G}^2}{\sigma_\mathcal{P}^2}\right)   \right]$.
\end{theorem}
\vspace{0.2in}
\begin{proof}
See~\cite{waterslide}. There is a better bound in~\cite{waterslide} as well. This bound is presented here for ease of exposition. 
\end{proof}
Observe that the required value of $n$ increases as $\pe$ decreases. Taking log on both sides of~\eqref{eq:peip}, it is evident that for small $\pe$, the term $nD(\sigma_\mathcal{G}^2||\sigma_\mathcal{P}^2)$ dominates the other terms in the RHS. For small $\pe$, $\sigma_\mathcal{G}^2$ can be taken close to $\sigma_\mathcal{G}^{*2}$ that satisfies $C_{\sigma_\mathcal{G}^{*2}}(P_T)=R$. Neglecting the other two terms, we get 
\begin{eqnarray}
\label{eq:approxbound}
n&\gtrsim& \frac{\log(1/\pe)}{D(\sigma_\mathcal{G}^{*2}||\sigma_\mathcal{P}^2)}.
\end{eqnarray}

\section{Minimization of total energy by optimizing over the rate and transmit power.}
\label{sec:minenergy}
Consider the total energy spent in transmission. For transmitting $k$
bits at rate $R$, the number of channel uses is $m = k/R$. If each
transmission has power $\xi_T P_T$, the total energy used in
the transmissions is  $\xi_T P_T m$. 

\begin{figure}[htb]
\begin{center}
\includegraphics[scale=0.45]{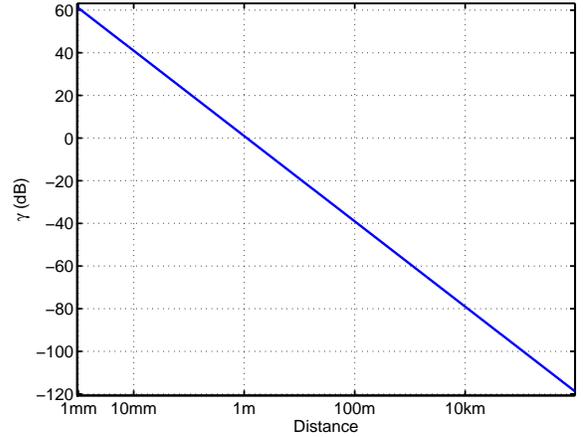}
\caption{The plot shows the behavior of $\gamma$ with distance $d$ for path loss $\frac{1}{\xi_T}=\frac{1}{d^2}$ for $d>0.1$mm (and path loss $1$ for smaller $d$). $E_{node}$ is $1$pJ, $\alpha = 4$, $\xi_D=1$, and $\sigma_\mathcal{P}^2=4\times 10^{-21}$J. The energy per bit is normalized by $\sigma_\mathcal{P}^2$.}
\label{fig:gammavsdistance}
\end{center}
\end{figure}

At the decoder, let the number of iterations be $l$. Assume that each
node consumes $E_{node}$ joules of energy in each iteration. The
number of computational nodes can be lower bounded by $m$, the number 
of received channel outputs, and also by $k$, the number of bits to be decoded. We lower bound by the maximum of the two\footnote{A lower bound of $m+k$ would not allow for node sharing between the set of observation nodes and the message nodes.}
\begin{equation}
E_{dec}\geq E_{node}\times \max\{k,m\}\times l.
\end{equation} 
There is no lower bound on the encoding complexity and so the
encoding is considered free. For $m$ transmissions with average power $P_T$, we require $mP_T$ joules of energy. This results in the following bound for
the weighted total energy\footnote{The parameters $\xi_T$ and $\xi_D$ are weights assigned to the transmit and the decoding energy respectively. $\xi_T$ depends on the path-loss across the channel. $\xi_D$ indicates the relative importance of decoding energy. For example, if the energy use at the decoder is severely constrained, $\xi_D$ would be large.}
\begin{equation}
E_{total}\geq \xi_T mP_T + \xi_D E_{node} \max\{k,m\}\times l.
\end{equation} 
Using $l \geq \frac{\log(n)}{\log(\alpha)}$,
\begin{eqnarray}
\label{eq:ptotal}
\nonumber E_{total} &\geq & m\xi_T P_T +\frac{\xi_D E_{node}\max\{k,m\} \log(n)}{\log(\alpha)}\\
& \propto & \frac{mP_T}{\sigma_\mathcal{P}^2} + \gamma\max\{k,m\} \log(n),
\label{eq:totavg}
\end{eqnarray} 
where $\gamma = \frac{\xi_D E_{node}}{\sigma_\mathcal{P}^2 \xi_T \log(\alpha)}$
is a constant that summarizes all the technological and environmental
terms. The expression in~\eqref{eq:ptotal} gives the normalized total energy, normalized by the noise variance $\sigma_\mathcal{P}^2$. Figure~\ref{fig:gammavsdistance} provides example\footnote{The energy cost of one iteration at one node $E_{node}\approx 1$ pJ is arrived at by an optimistic extrapolation from the reported values in~\cite{manishAllerton07, HowardSchlegel}, thermal noise
  energy per sample $\sigma_\mathcal{P}^2 \approx 4 \times 10^{-21}$J from $kT$
  with $T$ around room temperature.} behavior of $\gamma$ with distance.  The neighborhood size $n$ itself can be lower bounded by
plugging the desired average probability of error into
Theorem~\ref{thm:basicAWGNbound}.

\begin{figure}[htb]
\begin{center}
\includegraphics[scale=0.45]{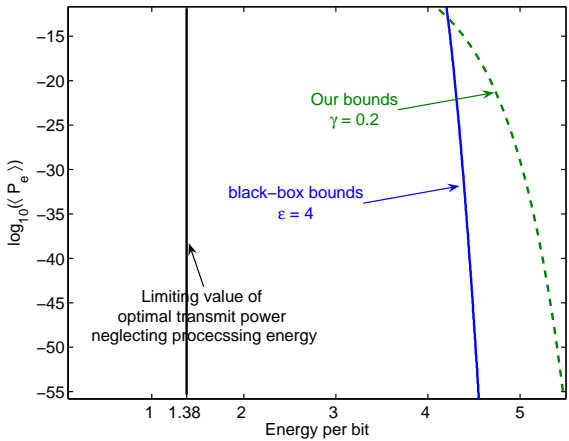}
\caption{The plot shows the comparison of lower bounds on the minimum normalized energy for $k=10,000$ bits. The `black-box bounds' plot is based on the model in~\cite{massaad}, where the details of the processor are ignored. Our bounds take into account the decoder structure as well. }
\label{fig:comparison}
\end{center}
\end{figure}

We thus obtain the following expression for the minimum normalized total energy,
\begin{eqnarray}
\label{eq:minenergy}
E_{\text{per bit}} = \min_{P_T,R}\frac{1}{R}\frac{P_T}{\sigma_\mathcal{P}^2} + \frac{1}{R} \gamma \max\left\{\frac{1}{R},1\right\}\log(n).
\end{eqnarray}
Observe that in~\eqref{eq:minenergy}, the decoding energy increases as the error probability decreases for constant transmit power and rate. This behavior is not reflected by using the model inspired from~\cite{massaad} for decoding energy. The bounds in~\cite{massaad} are for error probability converging to zero. To compare our bounds with the black-box model of~\cite{massaad}, in Appendix~\ref{app:nonzerope} we derive bounds for non-zero error probability based on the model in~\cite{massaad}. We plot the two bounds against each other in Figure~\ref{fig:comparison} for $k=10,000$ bits. 

We choose $\epsilon=4$, for which the total energy per bit for the black-box model equals the energy per bit for $\gamma=0.2$ for our bound for $\pe = 10^{-13}$. The figure shows that for $\pe$ smaller than this threshold, the model inspired from~\cite{massaad} underestimates the total energy. It is because this model treats the decoder as a black-box where $\epsilon$ does not change with error probability or rate. 

It is interesting to observe what values of $R$ optimize~\eqref{eq:minenergy}.   Under the small $\pe$ approximation in~\eqref{eq:approxbound}, we now heuristically argue that the optimal rate $R_{\mbox{opt}}$ should converge to $1$ as $\pe\rightarrow 0$.

Observe that for $R<1$,
\begin{eqnarray*}
E_{\mbox{per bit}} = \frac{P_T}{\sigma_\mathcal{P}^2 R}+\frac{\gamma}{R}\lo{n}\\
=  \frac{P_T}{\sigma_\mathcal{P}^2R} + \frac{\gamma}{R}\lo{\lo{\frac{1}{\pe}}}-\lo{{D(\sigma_\mathcal{G}^{*2}||\sigma_\mathcal{P}^2)}}
\end{eqnarray*}
As $\pe\rightarrow 0$, $n\rightarrow\infty$. Therefore, the decoding energy increases to infinity. Increasing the rate $R$ at the cost of increasing $P_T$ offsets the increasing decoding costs. However, for $R\geq 1$, 
\begin{equation}
E_{\mbox{per bit}} \gtrsim  \frac{P_T}{\sigma_\mathcal{P}^2R} + \gamma \lo{\frac{\lo{\frac{1}{\pe}}}{D(\sigma_\mathcal{G}^{*2}||\sigma_\mathcal{P}^2)}},
\end{equation}
which indicates  there is no advantage in increasing rate beyond $R=1$, since it no longer decreases the decoding energy.

\begin{figure}[htb]
\begin{center}
\includegraphics[scale=0.45]{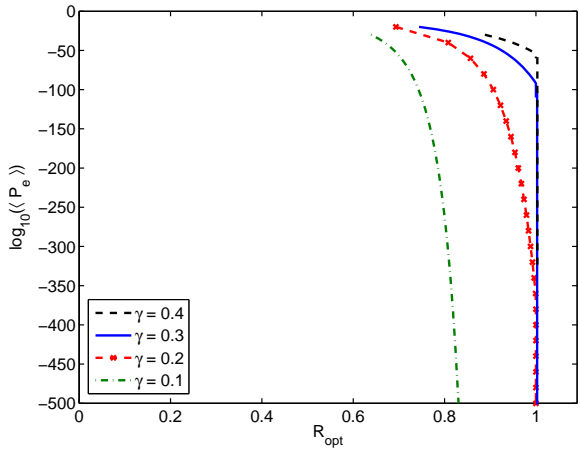}
\caption{Optimal value of rate vs error probability: As $\pe$ converges to $0$, the optimizing rate converges extremely slowly to $1$.}
\label{fig:roptvspe}
\end{center}
\end{figure}

Evidently, for finite $\pe$, there exists an optimal rate $R_{\text{opt}}>0$ that minimizes the combined energy consumed. Using numerical evaluation of the bound~\eqref{eq:minenergy}, we plot the behavior of the optimal rate with $\pe$ in Figure~\ref{fig:roptvspe}. The plots demonstrate that the optimal rate indeed converges to $1$.

\begin{figure}[htb]
\begin{center}
\includegraphics[scale=0.45]{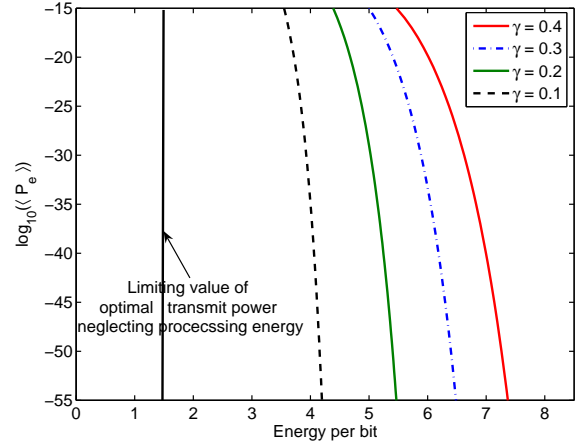}
\caption{The plot shows the behavior of lower bound on the normalized sum energy with $\pe$ for various values of $\gamma$. The sum energy goes to infinity as $\pe\rightarrow 0$.}
\label{fig:PeVsEnergyVarious}
\end{center}
\end{figure}

\begin{figure}[htb]
\begin{center}
\includegraphics[scale=0.45]{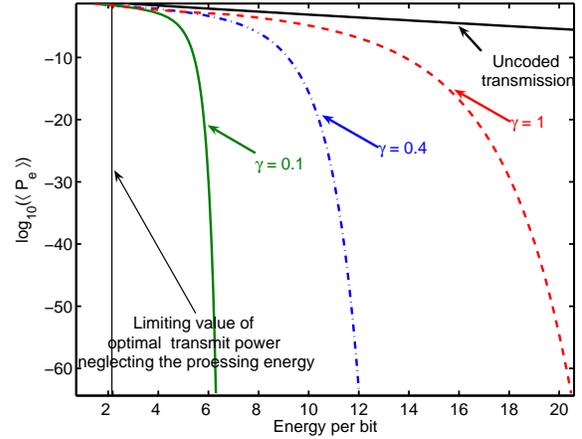}
\caption{The plot shows the behavior of lower bound on normalized sum energy with $\pe$ for various values of $\gamma$ for a BSC arising from performing hard-decision on BPSK symbols transmitted over an AWGN channel. The optimizing rate converges to 1 as $\pe\rightarrow 0$. Even so, this plot shows that the optimal strategy is not uncoded transmission at low $\pe$ since coded transmission outperforms uncoded transmission at small $\pe$.}
\label{fig:minenergyvspe}
\end{center}
\end{figure}

Figure~\ref{fig:PeVsEnergyVarious} shows the behavior of our lower bound on sum energy with $\pe$ for various values of $\gamma$. Figure~\ref{fig:minenergyvspe} shows that similar behavior also holds for a BSC arising from performing hard-decision on BPSK symbols transmitted over an AWGN channel.   The optimal rate for this channel also converges to $1$ as $\pe\rightarrow 0$. Due to lack of space, we omit the plots.


\section{Discussions and Conclusions}
\label{sec:conclusions}
In this work, we derived lower bounds on the combined transmission and decoding energy for iterative decoding with unconstrained rates. It is important to note that these are lower bounds, and the actual energy consumption would only be higher. An interesting feature of the our bounds is that the optimizing rate for green codes is bounded away from zero, and, in fact, converges to $1$ as the error probability converges to zero. This is qualitatively different from a pure black-box modeling of the decoding process, where energy consumption is independent of the desired error probability and the rate. In that case, as observed in~\cite{massaad}, the optimal rate is a constant that can be greater than $1$. 

For an AWGN channel, the value $1$ for optimal rate is a result of a bit-wise representation of the information at the decoder. If, however, the message nodes represent the information in base $M$ then the optimizing rate would converge to $\lo{M}$.

For the BSC arising from performing hard-decision on BPSK symbols transmitted over an AWGN channel, the optimal rate still converges to $1$. The rate is  upper bounded by $1$ because the channel has binary input alphabet, and thus this case might seem somewhat uninteresting. However, uncoded transmission over BSC also corresponds to rate $1$, which might falsely suggest that uncoded transmission is asymptotically optimal for minimizing the total energy. We observe that despite the optimal rate approaching $1$, coded transmission attains the same error probability with much smaller total energy than uncoded transmission. 

We note that the total energy per-bit required to communicate at arbitrarily low error probability increases to infinity for the message passing decoder. This is in  contrast to the classical information-theoretic result for transmit power, which shows that the transmit power is bounded even as $\pe\rightarrow 0$. Based on results in~\cite{waterslide}, the total energy per bit increases to infinity for most known codes and decoding algorithms. It would be interesting to extend this result to all possible codes and decoding algorithms. An approach based on laws of physics is suggested in~\cite{waterslide} for the fixed rate problem. The approach might yield results here as well. 

\appendices
\section{Bounds in~\cite{massaad} for non-zero error probability}
\label{app:nonzerope}
Observe that the results in~\cite{massaad} are for $\pe\rightarrow 0$ and infinitely many information bits. Parallel to our analysis for message passing decoding, in this appendix, we build on the analysis in~\cite{massaad} to derive bounds on the minimum energy required for communicating with a non-zero error probability $\pe$ and finite information bits. 

Assume $k$ bits are to be transmitted across the channel, with desired error probability $\pe$. In~\cite{massaad}, the authors maximize the information bits communicated under a total energy constraint. Turning around the problem in~\cite{massaad}, we can instead minimize the total energy consumed given the number of bits transmitted. Now we can add an error probability constraint to the bits transmitted. Assume that a block code is used to communicate across the channel. The corresponding error exponent is bounded by the sphere-packing bound~\cite{ShannonSpherePacking}. Assuming optimistically that the code actually achieves the sphere-packing bound in the exponent,
\begin{eqnarray*}
\pe \leq P_{e,block}\approx e^{-mE_{sp}(P_T,R)}
\end{eqnarray*}
where $E_{sp}(P_T,R)$ is the sphere-packing bound at rate $R$ and transmit power $P_T$. The objective, therefore, is
\begin{eqnarray}
\label{eq:massaadprob}
\nonumber&\underset{P_T,m}{\min}& m\times (P_T+\epsilon)\\
&\text{subject to}& m\times E_{sp}\left(P_T,\frac{k}{m}\right)=\lon{\frac{1}{\pe}}.
\end{eqnarray}

\bibliographystyle{IEEEtran}
\bibliography{IEEEabrv,MyMainBibliography}

\end{document}